\documentclass[
amsmath,amssymb,
aps,prd,
groupedaddress]{revtex4-2}

\usepackage{graphicx}
\usepackage{bm,array,slashed,braket,tensor}
\usepackage{hyperref}
  \hypersetup{colorlinks=true} 
\usepackage{comment} 
\newcommand{\1}{\mbox{1}\hspace{-0.25em}\mbox{l}} 
\newcommand{\In}{\mathrm{in}} %
\newcommand{\Out}{\mathrm{out}} %

\begin{document}

\title{Dynamically assisted pair production in subcritical potential step and particle--anti-particle interpretations}
\author{Makoto Ochiai}
\email[]{ochiai36@akane.waseda.jp}
\affiliation{Waseda Research Institute for Science and Engineering, Waseda University, 169-8555 Tokyo, Japan}
\date{October 20, 2023}

\begin{abstract}
  Particle--anti-particle interpretation under spatially inhomogeneous external fields within the framework of quantum field theory is a nontrivial problem. In this paper, we focus on the two interpretations established in [Phys. Rev. D {\bf 93}, 045002 (2016)] and [Prog. Theor. Exp. Phys. {\bf 2022}, 073B02 (2022)], both of which give consistent results of vacuum instability and pair production. To shed light on their differences, a pair production under a potential step assisted by a weak and oscillating electric field is discussed. It is shown that the potential step and the oscillating field, each insufficient for vacuum decay, can produce pairs when combined. In addition, the two pictures give rise to quantitative differences in the number of created pairs at the second-order perturbation of the oscillating field. It might provide a clue to investigate the correct particle--anti-particle interpretation by comparing the result with numerical simulations or experiments. 
\end{abstract}


\maketitle

\section{\label{Sec1}Introduction}

The particle--anti-particle pair production from the vacuum under external fields has been discussed in a wide variety of areas such as particle physics, nuclear physics, cosmology, and astrophysics~\cite{greinerQuantumElectrodynamicsStrong1985, ruffiniElectronPositronPairs2010}. In particular, in the case of strong electric fields, it is known as the Schwinger effect. It was first predicted by Sauter's observation~\cite{sauterUberVerhaltenElektrons1931, *sauterKleinschenParadoxon1932} of exact solutions of the Dirac equation under a constant homogeneous electric field, in association with the Klein paradox~\cite{kleinReflexionElektronenPotentialsprung1929}. In later years, many physicists, including Heisenberg, Euler, and Schwinger~\cite{heisenbergConsequencesDiracTheory2006, schwingerGaugeInvarianceVacuum1951} have revealed its non-perturbative aspects in quantum field theory. Nowadays, it is naively understood as a kind of dielectric breakdown of a quantum vacuum filled with virtual pairs of particles and anti-particles. The pair production accompanied by the vacuum instability is exponentially suppressed in many cases, and its direct detection needs electric fields with incredibly high intensity, given by the Schwinger limit. In recent years, a situation in which a strong electric field is superimposed on a weak and oscillating electromagnetic field has attracted physicists' attention. Perturbative contribution is combined with the non-perturbative one to yield a new interplay effect, drastically enhancing the pair production. The process is called the dynamically assisted Schwinger effect~\cite{schutzholdDynamicallyAssistedSchwinger2008}, and its experimental verification with intense laser facilities is expected to be within reach in the near future~\cite{fedotovAdvancesQEDIntense2023a}.

One of the simplest and most powerful tools when discussing the pair-creating phenomena is the canonical quantization of fields using mode functions, known as the Furry picture~\cite{fradkinFurryPictureQuantum1981, fradkinQuantumElectrodynamicsUnstable1991}. One solves the Dirac equation under external fields to expand a field operator on the basis of a complete orthonormal set of its solutions (mode functions). Creation/annihilation operators are introduced as the expansion coefficients. The key ingredient is how to define physically appropriate particles and anti-particles under the influence of external fields. The particle--anti-particle concept can be well-posed when a strong electric field depends only on time and switches off (asymptotically) in the distant past and future $t = \pm\infty$. In this case, the Dirac equation reduces to a first-order linear ordinary differential equation in terms of time. Its non-stationary solutions that asymptotically behave as single-mode plane waves in the distant past/future are called ``in/out'' mode functions, providing a physical vacuum and corresponding particle and anti-particle states, which can be interpreted in the distant past/future. Thus, various transition amplitudes between the two asymptotic times and expectation values of physical quantities can be calculated. Previous studies have evaluated vacuum persistence probability, the pair production number, etc., under time-dependent strong electric fields, using exact solutions~\cite{gavrilovVacuumInstabilityExternal1996, *adornoParticleCreationPeak2016, *adornoExactlySolvableCases2017a, *breevVacuumInstabilityTimedependent2021} or approximated solutions by the WKB method~\cite{kimImprovedApproximationsFermion2007, *kimEffectiveActionQED2008, hebenstreitMomentumSignaturesSchwinger2009,  *dumluStokesPhenomenonSchwinger2010, *dumluInterferenceEffectsSchwinger2011, dabrowskiTimeDependenceAdiabatic2016, tayaExactWKBAnalysis2021a}. Some studies~\cite{torgrimssonDynamicallyAssistedSauterSchwinger2017, tayaFranzKeldyshEffectStrongfield2019a} have discussed the dynamically assisted Schwinger effect by incorporating the perturbation theory into the mode functions method, where strong electric fields are treated non-perturbatively, while weak fields perturbatively (Furry-picture perturbation theory).

Spatial inhomogeneity in the gauge backgrounds makes the discussion much more complicated than the time-dependent case because one must solve the Dirac equation as a partial differential equation. Even if we assume that the external fields are formally time-independent by neglecting their switching-on/off effect, the equation reduces to a stationary Dirac equation, and its stationary solutions never satisfy the boundary conditions for the ``in/out'' mode functions. Thus, defining the physical vacua and particle--anti-particle pictures in the spatially inhomogeneous backgrounds is nontrivial. The problem is related to the Klein paradox or Klein tunneling~\cite{calogeracosKLEINTUNNELLINGKLEIN1999, *dombeySeventyYearsKlein1999, *calogeracosHistoryPhysicsKlein1999}, where scattering of a relativistic particle off a high potential step is considered. For the numerical simulations of these phenomena, see \cite{krekoraKleinParadoxSpatial2004, *krekoraKleinParadoxSpinresolved2005, *chengIntroductoryReviewQuantum2010, jiangElectronpositronPairCreation2011, *jiangPairCreationEnhancement2012, aleksandrovElectronpositronPairProduction2016, *aleksandrovMomentumDistributionParticles2017, *aleksandrovPairProductionTemporally2020a, ababekriEffectsFiniteSpatial2019a, schneiderDynamicallyAssistedSauterSchwinger2016, torgrimssonSauterSchwingerPairCreation2018, villalba-chavezSignaturesSchwingerMechanism2019, renDynamicallyAssistedSchwinger2023, hebenstreitSchwingerEffectInhomogeneous2011, kohlfurstElectronpositronPairProduction2015}.

We first remark on Nikishov's work from the viewpoint of gauge invariance in relativistic quantum mechanics~\cite{nikishovPairProductionConstant1969, *nikishovBarrierScatteringField1970}, where Green's functions under the constant homogeneous electric field are introduced by using mode functions in two gauges. This electric field is brought not only from a time-dependent vector potential (a temporal gauge) but also from a position-dependent scalar potential (a spatial gauge). A vacuum decay rate and a pair-production number are evaluated in both gauges to confirm the coincidence. The mode functions in a spatial gauge are characterized by the boundary conditions at spatial infinity instead of those in the distant past and future. The criteria for the mode functions are based on the assumption that particles and anti-particles, if appropriately defined, should be in the spatial infinity at initial and final times. Nikishov applied the criteria to scalar potentials with one-dimensional spatial inhomogeneity, which cannot be deformed into temporal gauges, such as a hyperbolic tangent potential (Sauter potential) and a step potential~\cite{nikishovScatteringPairProduction2004}. The vacuum decay rate for the Sauter potential is in good agreement with the one calculated using a path-integral based technique called the worldline method~\cite{dunneWorldlineInstantonsPair2005, *dunneWorldlineInstantonsFluctuation2006, giesPairProductionInhomogeneous2005, schneiderWorldlineInstantonsSauterSchwinger2019}. Gavrilov and Gitman incorporate Nikishov's particle--anti-particle picture into a framework of quantum field theory to investigate various physical quantities such as an electric current, an energy-momentum tensor, etc., in a vacuum state or a one-particle states~\cite{gavrilovQuantizationChargedFields2016}. They confirm that the results are consistent with the conventional hole picture, where particles in the so-called Dirac sea spontaneously tunnel into the positive-frequency area, yielding a current of particle--anti-particle pairs.

There is another attempt to develop the quantum field theory under the potential step on the basis of different characterizations of asymptotic states~\cite{nakazatoUnstableVacuumFermion2022}. In this work, one does not choose ``in/out'' mode functions from the start; instead, one tries to observe the asymptotic behavior of the field operator and determine the mode functions in an actual calculation. The basic idea is that asymptotic creation/annihilation operators equipped with appropriate particle--anti-particle interpretation are accompanied by monochromatic plane waves included in the field operator at asymptotic times. Thus, one calculates Dirac inner products of plane waves and the field operator in the limit $t \to \pm \infty$ to obtain those creation/annihilation operators. To do this, the field operator is expanded on the basis of a particular complete orthonormal set with formal creation/annihilation operators and quantized. Here, the formal operators just play the role of parameters connecting the physical creation/annihilation operators. Eliminating the parameters leads to the so-called Bogoliubov transformation, which precisely agrees with Gavrilov and Gitman's formula~\cite{gavrilovQuantizationChargedFields2016}. This result is also consistent with the other relevant works~\cite{hansenKleinParadoxIts1981, kimEffectiveActionQED2010, chervyakovExactPairProduction2009, *chervyakovElectronPositronPair2018, gavrilovScatteringPairCreation2016, evansParticleProductionFinite2021}.

The two frameworks \cite{gavrilovQuantizationChargedFields2016} and \cite{nakazatoUnstableVacuumFermion2022} are, in fact, partially inconsistent with each other in terms of the choice of the mode functions, and thus, particle--anti-particle interpretation. However, they give the same Bogoliubov transformation, and the inconsistency does not cause any quantitative differences in the discussion of vacuum decay and pair production under the stationary external fields with spatial inhomogeneities. In this paper, we superimpose a fluctuating field on the one-dimensional potential step and evaluate the dynamically assisted pair production. It is shown that the two frameworks yield different particle numbers at the second-order perturbation of the fluctuating field. Although there is no guarantee that either of them characterizes the correct particle--anti-particle picture, the result implies that the dynamical assistance effect might exhibit dependence on the definition of particles and anti-particles in the pair-creation phenomenon.

The paper is organized as follows: in the next section, we review the two frameworks adopted in \cite{nakazatoUnstableVacuumFermion2022} and \cite{gavrilovQuantizationChargedFields2016} under a position-dependent strong electric field, which is brought by a scalar potential. We call them the pictures (A) and (B), respectively, and see their differences in the context of quantum field theory. In Sec.~\ref{Sec3}, a weak and oscillating electric field, given as a vector potential, is incorporated as a perturbation. The particle numbers created from the vacuum are calculated in each framework and displayed in the subsequent sections: in Sec.~\ref{Sec4}, several features in the momentum distribution of the particle number with their underlying physics are discussed, and in Sec.~\ref{Sec5}, the dependence on the different particle--anti-particle pictures on the results is shown. Sec.~\ref{Sec6} is devoted to the conclusion and future work. Comments on spinors and mode functions are added in Appendices~\ref{AppendA} and \ref{AppendB}, respectively.

\section{\label{Sec2}Two frameworks with different particle and anti-particle interpretations}

First, the field-theoretical frameworks under a strong electric field, (A) in \cite{nakazatoUnstableVacuumFermion2022} and (B) in \cite{gavrilovQuantizationChargedFields2016}, are reviewed. Natural units $\hbar = c = 1$ are adopted throughout the paper. A field of relativistic fermion with mass $m$ under the influence of a step potential alone is evolved by the Dirac equation 
\begin{align} 
  [i\gamma^0 (\partial_t - ieA_0 (z)) + i\gamma^3 \partial_z - m] \Psi^{(0)} = 0, \label{eq::DiracEq_Psi(0)} 
\end{align} 
where $\gamma^0, \gamma^3$ are $4\times 4$ gamma matrices and $e > 0$ is a magnitude of an electron's charge. Here, the dependence on spatial coordinates other than $z$ is neglected for simplicity. The scalar potential $A_0 (z)$ stands for a one-dimensional step potential along $z$-direction: denoting $\theta (z)$ as a step function, 
\begin{align} 
  V(z) = -eA_0 (z) = V_0 \theta (z), \label{eq::V(z)} 
\end{align} 
with the potential height $V_0$. We consider exclusively the subcritical case $V_0 < 2m$, where the potential does not induce vacuum instability and pair production. The superscript on the field implies that the solution of \eqref{eq::V(z)} will be used as the unperturbed one when an oscillating field is added as a perturbation. The step potential gives a time-independent electric field localized at $z = 0$. The equation of motion \eqref{eq::DiracEq_Psi(0)} can be written as the Schr\"{o}dinger-like equation 
\begin{align} 
  i\partial_t \Psi^{(0)} = H\Psi^{(0)} \label{eq::Schroedinger-likeEq} 
\end{align} 
with 
\begin{align} 
  H = -i\alpha_z \partial_z + \beta m + V(z), \label{eq::DiracH} 
\end{align} 
($\alpha_z = \gamma^0 \gamma^3, \beta = \gamma^0$) and the mode functions adopted in (A) and (B) are both stationary solutions of \eqref{eq::Schroedinger-likeEq}. Their explicit forms are described in the next two subsections.

The energy spectra of the Dirac Hamiltonian \eqref{eq::DiracH} are classified into four regions: for an energy eigenvalue $E$, 
\begin{itemize} 
  \item[(i)] $E > V_0 + m$, 
  \item[(ii)] $m < E \leq V_0 + m$, 
  \item[(iii)] $-m \leq E < V_0 - m$, and 
  \item[(iv)] $E < -m$. 
\end{itemize} 
Since the eigenfunctions in regions (ii) and (iii) are uniquely determined due to a mass gap, the mode functions in (A) and (B) are the same up to their normalization factors. In the other energy regions, however, they are doubly degenerated, and thus, there remains a possibility of choosing different mode functions and the different particle--anti-particle pictures (A) and (B).

\subsection{\label{Sec2.1}Particle--anti-particle picture (A)}

\begin{figure}[htbp]\centering 
  \includegraphics[width=0.75\linewidth]{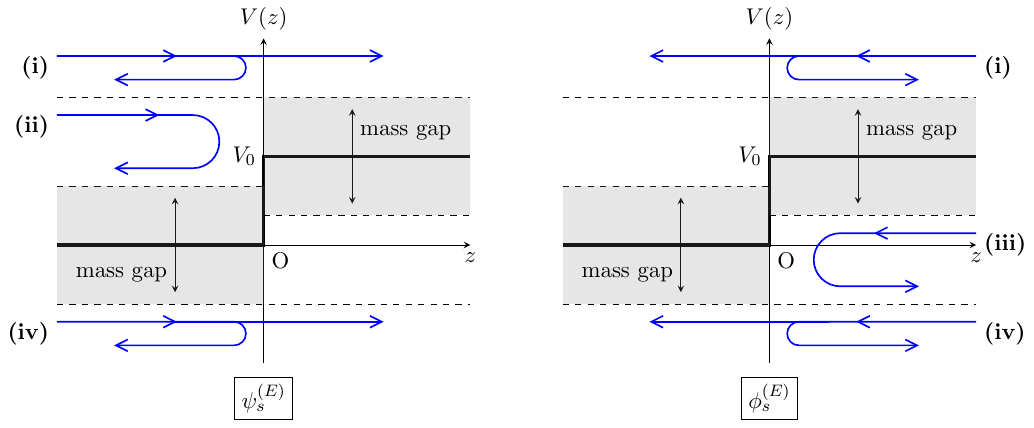} 
  \caption{\label{fig::psi_phi}The scattering wave functions belonging to the energy regions (i)--(iv) under the subcritical potential step (with height $V_0 < 2m$). Blue arrows represent directions of incident, reflected, and transmitted waves in the scattering wave functions. Gray-shaded regions are a mass gap, where oscillating solutions do not exist. The left half corresponds to the left-incident case, while the right half corresponds to the right-incident case. } 
\end{figure}

In \cite{nakazatoUnstableVacuumFermion2022}, scattering wave functions are adopted as an expansion basis of the Dirac field $\Psi^{(0)}$. A left-incident scattering wave function $\psi_s^{(E)}$ in an energy eigenvalue $E \in \text{(i)}$ and a spin $s$, is composed of the incident and transmitted waves moving to the right (positive $z$-direction) and a reflected wave moving to the left (negative $z$-direction). The momentum of these waves is determined by the energy-momentum relation $E = E_p = V_0 + E_q$ ($E_p = \sqrt{p^2 + m^2}$), where $p$ and $q$ denote the magnitudes of momenta of the initial and transmitted waves, respectively. $\psi_s^{(E)}$ is expressed by using a Dirac spinor of positive frequency $u$ (see Appendix~\ref{AppendA}) as 
\begin{align} 
  \psi_s^{(E)} (z, t) = \frac{1}{\sqrt{2\pi}} \sqrt{\frac{m}{E}} e^{-iEt} \Bigl\{ \theta (-z) \bigl[ u(p, s) e^{ipz} + R_\psi (p) u(-p, s) e^{-ipz} \bigr] + \theta (z) T_\psi (p) u(q, s) e^{iqz} \Bigr\}, \label{eq::psi_(i)} 
\end{align} 
with reflection and transmission coefficients 
\begin{align} 
  R_\psi (p) = \frac{\sqrt{\frac{E - V_0 + m}{E + m}} - \sqrt{\frac{E - V_0 - m}{E - m}}}{\sqrt{\frac{E - V_0 + m}{E + m}} + \sqrt{\frac{E - V_0 - m}{E - m}}}, \quad T_\psi (p) = \frac{2}{\sqrt{\frac{E - V_0 + m}{E + m}} + \sqrt{\frac{E - V_0 - m}{E - m}}}. \label{eq::Rpsi_Tpsi} 
\end{align} 
They are determined by the continuity condition for the scattering wave function at the discontinuous point of the potential $z = 0$. The continuity of a vector current along $z$-axis $j = \bar{\psi}_s^{(E)} \gamma^3 \psi_s^{(E)} = \psi_s^{(E) \dagger} \alpha_z \psi_s^{(E)}$ at $z = 0$ gives the so-called probability conservation:  
\begin{align} 
  P_{\rm refl} + P_{\rm trans} = |R_\psi (p)|^2 + \frac{q}{p} |T_\psi (p)|^2 = 1, \label{eq::prob_conserv} 
\end{align} 
where the reflection and transmission probabilities are defined as $P_{\rm refl} = |j_{\rm refl}/j_{\rm inc}|, P_{\rm trans} = |j_{\rm trans}/j_{\rm inc}|$ with incident, reflected and transmitted currents $j_{\rm inc}, j_{\rm refl}, j_{\rm trans}$. The above function \eqref{eq::psi_(i)} describes an over-the-barrier scattering of a positive-frequency wave. Note that the directions of each current coincide with those of phase velocity and group velocity. The energy-momentum relation also holds for the right-incident case in the same energy region. The right-incident scattering wave function $\phi_s^{(E)}$ is expressed as 
\begin{align} 
  \phi_s^{(E)} (z, t) = \frac{1}{\sqrt{2\pi}} \sqrt{\frac{m}{E - V_0}} e^{-iEt} \Bigl\{ \theta (-z) T_\phi (q) u(-p, s) e^{-ipz} + \theta (z) \big[ u(-q, s) e^{-iqz} + R_\phi (q) u(q, s) e^{iqz} \bigr] \Bigr\}, \label{eq::phi_(i)} 
\end{align} 
where $-p$ refers to the transmitted wave and $-q$ to the incident wave. The reflection and transmission coefficients are written as 
\begin{align} 
  R_\phi (q) = \frac{\sqrt{\frac{E + m}{E - V_0 + m}} - \sqrt{\frac{E - m}{E - V_0 - m}}}{\sqrt{\frac{E + m}{E - V_0 + m}} + \sqrt{\frac{E - m}{E - V_0 - m}}}, \quad T_\phi (q) = \frac{2}{\sqrt{\frac{E + m}{E - V_0 + m}} + \sqrt{\frac{E - m}{E - V_0 - m}}}, \label{eq::Rphi_Tphi} 
\end{align} 
which are related to those for the left-incident case through the reciprocal relations: 
\begin{align} 
  R_\phi (q) = -R_\psi (p), \quad T_\phi (q) = \frac{q}{p} T_\psi (p). \label{eq::reciprocity} 
\end{align} 
$\phi_s^{(E)}$ also describes an over-the-barrier scattering of a positive-frequency wave. For scattering wave functions in the other energy regions, see \cite{ochiaiCompletenessScatteringStates2018}. All the scattering behavior of waves described by the left- and right-incident scattering wave functions under the subcritical potential step \eqref{eq::V(z)} is displayed in Fig.~\ref{fig::psi_phi}. 
The incident waves in regions (ii) and (iii) penetrate the mass gap with exponential suppression and experience a total reflection. $\psi_s^{(E)}$ and $\phi_s^{(E)}$ in the region (iv) both exhibit the over-the-barrier scattering of negative-frequency waves. Note that for an overcritical potential step (with height $V_0 > 2m$), another energy region $m < E \leq V_0 - m$ (often called the Klein region) is present, where incident waves can transmit the barrier without exponential suppression. The tunneling differs from the usual one in non-relativistic quantum mechanics in that it occurs even in the rigid-wall limit $V_0 \to \infty$. This counter-intuitive effect is known as the Klein tunneling~\cite{sauterKleinschenParadoxon1932, calogeracosKLEINTUNNELLINGKLEIN1999, *dombeySeventyYearsKlein1999, *calogeracosHistoryPhysicsKlein1999}.

The scattering wave functions in every region (i)--(iv) are normalized as follows: 
\begin{align} 
  \int_{-\infty}^\infty dz \psi_s^{(E) \dagger} (z, t) \psi_{s'}^{(E')} (z, t) &= \theta (EE') \delta (p - p') \delta_{s, s'}, \label{eq::norm_psi} \\ 
  \int_{-\infty}^\infty dz \phi_s^{(E) \dagger} (z, t) \phi_{s'}^{(E')} (z, t) &= \theta ((E - V_0)(E' - V_0)) \delta (q - q') \delta_{s, s'}, \label{eq::norm_phi} 
\end{align} 
where $p$ and $q$ are the absolute values of momenta of the initial waves in $\psi_s^{(E)}$ and $\phi_s^{(E)}$, respectively. The step functions in the right-hand side of the above equations stand for the orthogonality between eigenfunctions belonging to different eigenvalues of a Hermitian operator \eqref{eq::DiracH}. The orthogonality between the degenerated eigenfunctions $\psi_s^{(E)}$ and $\phi_s^{(E)}$ seems nontrivial, but can be shown explicitly. Bound states do not exist, and all the scattering wave functions $\psi_s^{(E)}$ and $\phi_s^{(E)}$ form a complete set~\cite{ochiaiCompletenessScatteringStates2018, ruijsenaarsScatteringTheoryOnedimensional1977a}. The orthogonality and completeness are essential properties in the field quantization discussed below.

We expand the Dirac field $\Psi^{(0)}$ in the Heisenberg picture on the basis of the scattering wave functions as 
\begin{equation}\begin{split} 
  \Psi^{(0)} (z, t) &= \sum_s \int_0^\infty dp \Bigl( a_L (p, s) \psi_s^{(E_p)} (z, t) + b_L^\dagger (p, s) \psi_s^{(-E_p)} (z, t) \Bigr) \\ 
  &\quad + \sum_s \int_0^\infty dq \Bigl( a_R (q, s) \phi_s^{(V_0 + E_q)} (z, t) + b_R^\dagger (q, s) \phi_s^{(V_0 - E_q)} (z, t) \Bigr), \label{eq::field(A)} 
\end{split}\end{equation} 
with equal-time canonical anti-commutation relations: 
\begin{align} 
  \{ \Psi_\alpha^{(0)} (z, t), \Psi_\beta^{(0) \dagger} (z', t) \} = \delta (z - z') \delta_{\alpha, \beta}, 
\end{align} 
where $\alpha, \beta$ are spinor indices, and the other anti-commutators vanish. On the right-hand side of \eqref{eq::field(A)}, annihilation operators $a_L (p, s)$ and $a_R (q, s)$ are introduced as coefficients of $\psi_s^{(E_p)}$ in the energy regions (i)--(ii) and $\phi_s^{(V_0 + E_q)}$ in (i), whereas creation operators $b_L^\dagger (p, s)$ and $b_R^\dagger (q, s)$ in accordance to $\psi_s^{(-E_p)}$ in (iv) and $\phi_s^{(V_0 - E_q)}$ in (iii)--(iv). Because of the orthonormality and completeness of the expansion basis, the anti-commutation relations for the Dirac field are equivalent to those for the creation/annihilation operators: for example, for the annihilation operator $a_L (p, s)$, only the anti-commutator with itself is nonvanishing, giving a delta function 
\begin{align} 
  \{ a_L (p, s), a_L^\dagger (p', s') \} = \delta (p - p') \delta_{s, s'}. 
\end{align} 
It should be noted that the creation/annihilation operators are just the formal ones and are not equipped with particle--anti-particle interpretation. Aside from these operators, asymptotic creation/annihilation operators with physical meaning can be introduced by extracting wave modes characterizing particles and anti-particles from the field operator in the distant past and future $t = \pm \infty$. Particles and anti-particles in the area $z \neq 0$, if properly defined, are not subjected to an electric force from the potential step. Consider a scattering process; in the distant past and future, they should exist at the spatial infinity $|z| = \infty$ and be characterized by a monochromatic plane wave with positive and negative frequency, respectively. Since the field operator in \eqref{eq::field(A)} satisfies the equation of motion \eqref{eq::DiracEq_Psi(0)} at any time, the particle and anti-particle modes, along with the asymptotic creation/annihilation operators as their coefficients, should be included in the limit $t \to -\infty$ of the field operator. For instance, an ``in'' annihilation operator of particle at the left infinity $z = -\infty$ is defined as 
\begin{align} 
  a_{L, \In}^{(0)} (p, s) = \lim_{t \to -\infty} \int_{-\infty}^\infty dz u_{p, s}^\dagger (z, t) \Psi^{(0)} (z, t), \label{eq::a_Lin(0)_def} 
\end{align} 
where $u_{p, s} (z, t)$ is a positive-frequency plane wave on the left of the step 
\begin{align} 
  u_{p, s} (z, t) = \frac{1}{\sqrt{2\pi}} \sqrt{\frac{m}{E_p}} u(p, s) e^{-iE_p t + ipz}. 
\end{align} 
After substituting the field decomposition \eqref{eq::field(A)}, one has to evaluate the limit $t \to -\infty$ of the inner products of the plane wave and the scattering wave functions. Notice that almost all of them do not contribute to the limit because they indefinitely oscillate in time and thus vanish due to Riemann--Lebesgue's lemma. The candidates to remain are the singular terms without the oscillation factors, i.e., those composed of the plane wave and the scattering wave functions belonging to the energy eigenvalue $E_p$. The concrete calculation yields a simple expression of the annihilation operator of the ``in'' particle for $p > 0$, as 
\begin{align} 
  a_{L, \In}^{(0)} (p, s) = a_L (p, s). 
\end{align} 
The result provides the formal operator $a_L (p, s)$ with the physical meaning of a particle incoming from the left in the distant past. In other words, the left-incident scattering wave function $\psi_s^{(E_p)}$ plays a role of the ``in'' mode function in this framework (A). The same relations for the other creation/annihilation operators $b_L^\dagger (p, s), a_R (q, s), b_R^\dagger (q, s)$ and their counterparts $b_{L, \In}^\dagger (p, s), a_{R, \In} (q, s), b_{R, \In}^\dagger (q, s)$ are obtained. One finds that the corresponding scattering wave functions are the ``in'' mode functions in (A) which characterize a particle and an anti-particle incoming from the spatial infinity $|z| = \infty$. Thus, \eqref{eq::field(A)} can be understood as the field decomposition on the basis of the ``in'' mode functions in (A), where the subscripts ``in'' on the creation/annihilation operators should be added. Vacuum at the distant past, or ``in'' vacuum $\ket{0}_\In$, is defined as an eigenstate which is annihilated by any ``in'' annihilation operators: 
\begin{align} 
  a_{L, \In}^{(0)} (p, s) \ket{0}_\In = a_{R, \In}^{(0)} (q, s) \ket{0}_\In = b_{L, \In}^{(0)} (p, s) \ket{0}_\In = b_{R, \In}^{(0)} (q, s) \ket{0}_\In = 0. \label{eq::in_vac} 
\end{align} 
Any ``in'' states of the picture (A) are created by applying the creation operators on the ``in'' vacuum. Note that the right-hand side of \eqref{eq::a_Lin(0)_def} for $-p$ disappears. All information of the ``in'' particle and anti-particle approaching the center $z = 0$ are included in the field operator $\Psi^{(0)}$, but those going away from the center are not, at $t = -\infty$ limit.

The ``out'' creation/annihilation operators are introduced in the same way as the ``in'' creation/annihilation operators, but in the opposite limit $t \to \infty$. The field operator in the distant future $t = \infty$ contains the information of an outgoing particle and anti-particle away from the center. The explicit forms of the ``out'' creation/annihilation operators are obtained as a sum of the ``in'' creation/annihilation operators, satisfying the canonical anti-commutation relations; see Appendix~\ref{AppendB}. It should be mentioned that the ``in/out'' transformations hold among the asymptotic creation/annihilation operators in the same energy eigenvalue. It implies that particle--anti-particle mode mixing occurs only in the Klein region $m < E \leq V_0 - m$, where the positive-frequency spectrum of $\psi_s^{(E)}$ and the negative-frequency spectrum of $\phi_s^{(E)}$ overlap. In this case, the ``in'' vacuum becomes unstable to induce spontaneous pair production. The overcritical potential step supplies energy $V_0$ to a virtual particle--anti-particle pair by 
\begin{align} 
  E_p + E_q = V_0, \label{eq::EnergyBalance_step} 
\end{align} 
making it a real pair.  Note that ``out'' mode functions in (A), which are unnecessary in the above framework, are derived from the ``in/out'' transformation among the asymptotic creation/annihilation operators.

\subsection{\label{Sec2.2}Particle--anti-particle picture (B)}

\begin{figure}[htbp]\centering 
  \includegraphics[width=0.75\linewidth]{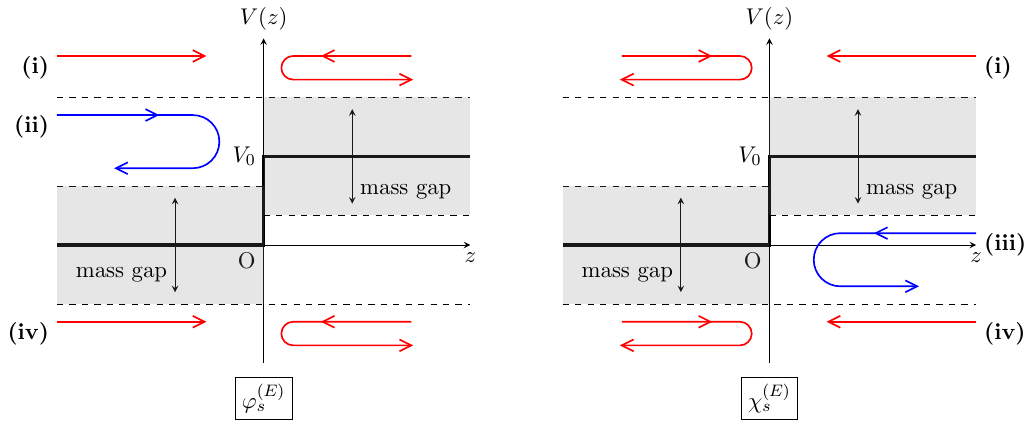} 
  \caption{\label{fig::varphi_chi}The ``in'' mode functions (B) in the energy regions (i)--(iv) under the same potential step. The left half of the figures shows $\varphi_s^{(E)}$, while the right half shows $\chi_s^{(E)}$. The directions of plane waves included in the mode functions are expressed in blue or red arrows. The blue arrows in the regions (ii) and (iii) are the same as those in Fig.~\ref{fig::psi_phi}.} 
\end{figure}

In the other particle--anti-particle picture (B), two complete orthonormal sets are prepared in the field decomposition. Although the previous work \cite{gavrilovQuantizationChargedFields2016} treated a scalar potential with more general configurations, we restrict to the case of the step potential for comparison with the picture (A). Two degenerated stationary solutions of the Dirac equation $\varphi_s^{(E)}$ and $\chi_s^{(E)}$ in the energy region (i) are expressed as 
\begin{align} 
  \varphi_s^{(E)} (z, t) &= \frac{1}{\sqrt{2\pi}} \sqrt{\frac{m}{E}} \sqrt{\frac{p}{q}} e^{-iEt} \Bigl\{ \theta (-z) T_\phi (q) u(p, s) e^{ipz} + \theta (z) \left[ u(q, s) e^{iqz} + R_\phi (q) u(-q, s) e^{-iqz} \right] \Bigr\}, \label{eq::varphi_(i)} \\ 
  \chi_s^{(E)} (z, t) &= \frac{1}{\sqrt{2\pi}} \sqrt{\frac{m}{E - V_0}} \sqrt{\frac{q}{p}} e^{-iEt} \Bigl\{ \theta (-z) \left[ u(-p, s) e^{-ipz} + R_\psi (p) u(p, s) e^{ipz} \right] + \theta (z) T_\psi (p) u(-q, s) e^{-iqz} \Bigr\}, \label{eq::chi_(i)} 
\end{align} 
where $R_\psi (p), T_\psi (p), R_\phi (q), T_\phi (q)$ are the same as in \eqref{eq::Rpsi_Tpsi} and \eqref{eq::Rphi_Tphi}. The above solutions are normalized by the conditions \eqref{eq::norm_psi} and \eqref{eq::norm_phi}, respectively. The normalizations are different from those in \cite{gavrilovQuantizationChargedFields2016}, where the delta functions of momenta in \eqref{eq::norm_psi} and \eqref{eq::norm_phi} are replaced with $\delta (E - E')$, but it does not change the particle--anti-particle notion provided the solutions satisfy the orthogonality and completeness relations. On the left of the step, $\varphi_s^{(E)}$ is composed of a single-mode plane wave moving to the right. Nikishov, and in later years, Gavrilov and Gitman, have interpreted this solution as particle mode ingoing from the left infinity in the distant past. The particle--anti-particle concept is brought about from the boundary conditions that the ``in'' mode functions behave asymptotically as single-mode plane waves approaching the center at the left infinity. The boundary conditions in terms of spatial infinity are in accordance with the case of time-dependent backgrounds as if the labels of space and time are interchanged. $\chi_s^{(E)}$ consists of a single-mode plane wave moving to the center on the right of the step, which is interpreted as particle mode incoming from the right infinity in the distant past. The behaviors of the ``in'' mode functions in (B) in every energy range are drawn in Fig.~\ref{fig::varphi_chi}. 
The ``in'' mode functions (B) in the regions (ii) and (iii), shown in blue arrows, are the same as the ``in'' mode functions (A), i.e., the scattering wave functions describing the total reflection. The mode functions (B) in the other regions, drawn in red arrows, differ from (A). The red arrows in the left (right) panel of Fig.~\ref{fig::varphi_chi} correspond to the blue arrows in the right (left) panel of Fig.~\ref{fig::psi_phi} whose directions are reversed. One finds that $\varphi_s^{(E)}$ and $\chi_s^{(E)}$ in \eqref{eq::varphi_(i)} and \eqref{eq::chi_(i)} coincide with the ``out'' mode functions (A) up to their normalizations (see Appendix~\ref{AppendB}). Thus, ``in/out'' separations and whether an (anti-)particle is on the left or right of the step are defined conversely. Note that the mode functions (A) and (B) in the same energy eigenvalue are related to each other by linear transformations: $\varphi_s^{(E)}$ and $\chi_s^{(E)}$ in the region (i) can be rewritten as a linear combination of $\psi_s^{(E)}$ and $\phi_s^{(E)}$, i.e., 
\begin{align} 
  \begin{pmatrix} \varphi_s^{(E)} (z, t) \\ \chi_s^{(E)} (z, t) \end{pmatrix} = \begin{pmatrix}  \sqrt{\frac{q}{p}} T_\psi (p) & \sqrt{\frac{p}{q} \frac{E_q}{E_p}} R_\phi (q) \\ \sqrt{\frac{q}{p} \frac{E_p}{E_q}} R_\psi (p) & \sqrt{\frac{p}{q}} T_\phi (q) \end{pmatrix} \begin{pmatrix} \psi_s^{(E)} (z, t) \\ \phi_s^{(E)} (z, t) \end{pmatrix}. \label{eq::lineartrans_modefunct} 
\end{align}

The field $\Psi^{(0)}$ is expanded on the basis of the ``in'' mode functions (B) as 
\begin{equation}\begin{split} 
  \Psi^{(0)} (z, t) &= \sum_s \int_0^\infty dp \Bigl( c_{L, \In}^{(0)} (p, s) \varphi_s^{(E_p)} (z, t) + d_{L, \In}^{(0) \dagger} (p, s) \varphi_s^{(-E_p)} (z, t) \Bigr) \\ 
  &\quad + \sum_s \int_0^\infty dq \Bigl( c_{R, \In}^{(0)} (q, s) \chi_s^{(V_0 + E_q)} (z, t) + d_{R, \In}^{(0) \dagger} (q, s) \chi_s^{(V_0 - E_q)} (z, t) \Bigr), \label{eq::field(B)} 
\end{split}\end{equation} 
where other ``in'' creation/annihilation operators are introduced as their coefficients. The anti-commutation relations for these operators are equivalent to the equal-time canonical anti-commutation relations for the field operator due to the completeness and orthonormality of the ``in'' mode functions (B). We denote the creation/annihilation operators with the subscript ``in'' since they have already been equipped with the physical meaning of particle--anti-particle by the above mode functions. One finds that ``in'' vacuum within the picture (B), introduced as a zero eigenstate for any annihilation operators in \eqref{eq::field(B)}, is the same as $\ket{0}_\In$ in \eqref{eq::in_vac}. It is confirmed by the relations among the ``in'' creation/annihilation operators in both pictures (A) and (B). To see this, let us write the ``in'' annihilation operator $c_{L, \In}^{(0)} (p, s)$, for example, as the inner products of the corresponding mode functions and the field operator: 
\begin{align} 
  c_{L, \In}^{(0)} (p, s) = \lim_{t \to -\infty} \int_{-\infty}^\infty dz \varphi_s^{(E_p) \dagger} (z, t) \Psi^{(0)} (z, t), \label{eq::c_Lin(0)_def} 
\end{align} 
where the limit $t \to -\infty$ is applied to the inner product in accordance with \eqref{eq::a_Lin(0)_def}, though it is unnecessary for an actual calculation. By substituting the field decomposition \eqref{eq::field(A)} and the linear transformation for $\varphi_s^{(E_p)}$ in \eqref{eq::lineartrans_modefunct}, the annihilation operator for the energy region (i) is expressed as 
\begin{align} 
  c_{L, \In}^{(0)} (p, s) = \sqrt{\frac{q}{p}} T_\psi (p) a_L (p, s) + \sqrt{\frac{p}{q} \frac{E_q}{E_p}} R_\phi (q) a_R (q, s). \label{eq::c_Lin(0)} 
\end{align} 
As shown in the ``in/out'' transformation among the annihilation operators in \eqref{eq::Bogoliubov_trans}, $c_{L, \In}^{(0)} (p, s)$ is proportional to the ``out'' annihilation operator $a_{R, \Out}^{(0)} (q, s)$.

The field operator $\Psi^{(0)}$ is also expanded on the basis of ``out'' mode functions (B) with ``out'' creation/annihilation operators equipped with particle--anti-particle notion and similar correspondence between (A) and (B) can be found. The ``out'' mode functions $\varphi_{s, \Out}^{(E)}$ and $\chi_{s, \Out}^{(E)}$ in the region (i) are identical to the ``in'' mode functions (A) up to their normalization factors: 
\begin{align} 
  \varphi_{s, \Out}^{(E)} (z, t) = \sqrt{\frac{p}{q} \frac{E_q}{E_p}} \phi_s^{(E)} (z, t), \quad \chi_{s, \Out}^{(E)} (z, t) = \sqrt{\frac{q}{p} \frac{E_p}{E_q}} \psi_s^{(E)} (z, t). \label{eq::modeft_AB} 
\end{align} 
They are normalized by \eqref{eq::norm_psi} and \eqref{eq::norm_phi}, respectively. The relations \eqref{eq::modeft_AB} are reflected in the creation/annihilation operators, such as 
\begin{align} 
  c_{L, \Out}^{(0)} (p, s) &= \lim_{t \to \infty} \int_{-\infty}^\infty dz \varphi_{s, \Out}^{(E_p) \dagger} (z, t) \Psi^{(0)} (z, t) = \sqrt{\frac{p}{q} \frac{E_q}{E_p}} a_R (q, s), \label{eq::c_Lout(0)} 
\end{align} 
where the limit $t \to \infty$ in the middle is actually unnecessary in the same way as \eqref{eq::c_Lin(0)_def}. It should be stressed again that particle--anti-particle mode mixing among the creation/annihilation operators occurs only in the Klein region. This is guaranteed by the orthogonality of the mode functions; see \eqref{eq::c_Lin(0)_def} and \eqref{eq::c_Lout(0)}.

Comments on the particle--anti-particle picture (B) for the overcritical case are in order. The ``in/out'' mode functions in the Klein region are the same as (A). The criteria for choosing the two sets of mode functions (B) are discussed by Gavrilov and Gitman~\cite{gavrilovQuantizationChargedFields2016} within the quantum field theory by calculating various physical quantities, such as an energy-momentum tensor in a one--(anti-)particle state. As discussed in the previous subsection, vacuum instability and pair production are attributed only to particle and anti-particle in the Klein region. Thus, quantitative differences between the two pictures do not emerge in the vacuum decay rate, the number of created pairs, etc.

\section{\label{Sec3}Furry-picture perturbation theory}

In the following discussion, a weak and oscillating electric field is incorporated into the quantum field theory developed in Sec.~\ref{Sec2} as a perturbation. We consider an oscillating gauge field with a single frequency $\omega$: 
\begin{align} 
  A_3 (z, t) = \frac{\mathcal{E}_z}{\omega} \sin (\omega t) e^{-(z/l)^2}, \label{eq::A_3(z,t)} 
\end{align} 
which is localized in space with a width $l$ along the same direction as the potential step. It gives a spatially localized and oscillating electric field with a maximum field strength $\mathcal{E}_z$, which is assumed to be much smaller than the Schwinger limit $m^2/e$. The time-evolution equation to be considered is now 
\begin{align} 
  [i\gamma^0 (\partial_t - ieA_0 (z)) + i\gamma^3 \partial_z - m] \Psi = -e\gamma^3 A_3 (z, t) \Psi, 
\end{align} 
and the right-hand side of it is treated as a perturbation term. The fermionic field $\Psi$ under the total external fields is expanded in a series of the perturbative gauge field $A_3$ up to the first order, as 
\begin{align} 
  \Psi (z, t) = \sqrt{Z} \Bigl[ \Psi^{(0)} (z, t) - e\int_{-\infty}^\infty dz' dt' S_{\rm ret} (z, t; z', t') \gamma^3 A_3 (z', t') \Psi^{(0)} (z', t') + \mathcal{O} \bigl( (A_3)^2 \bigr) \Bigr], \label{eq::Psi_series} 
\end{align} 
with an overall factor $\sqrt{Z}$ and the zeroth-order field $\Psi^{(0)}$ in the previous section. $S_{\rm ret}$ is a retarded Green's function, which obeys an equation 
\begin{align} 
  [i\gamma^0 (\partial_t - ieA_0 (z)) + i\gamma^3 \partial_z - m] S_{\rm ret} (z, t; z', t') = \delta (z - z') \delta (t - t'), \label{eq::EOM_Sret} 
\end{align} 
where the spinor indices are omitted. It is easy to show that the following form of $S_{\rm ret}$ 
\begin{align} 
  S_{\rm ret} (z, t; z', t') = -i\theta (t - t') \sum_{\epsilon, s = \pm} \biggl[ \int_0^\infty dp \, \psi_s^{(\epsilon E_p)} (z, t) \bar{\psi}_s^{(\epsilon E_p)} (z', t') + \int_0^\infty dq \, \phi_s^{(V_0 + \epsilon E_q)} (z, t) \bar{\phi}_s^{(V_0 + \epsilon E_q)} (z', t') \biggr] 
\end{align} 
satisfies \eqref{eq::EOM_Sret} along with the retarded boundary condition. Note that the right-hand side of $\Psi$ in \eqref{eq::Psi_series} is not a perturbation series of the coupling constant $e$ because the scalar potential $A_0$ with its coupling constant is incorporated non-perturbatively to the zeroth-order field and the retarded Green's function. The non-perturbative contributions from $A_0$ as well as the perturbative ones from $A_3$ are combined in the perturbation expansion \eqref{eq::Psi_series}, in which the interplay between the non-perturbative and perturbative mechanisms in \cite{torgrimssonDynamicallyAssistedSauterSchwinger2017, tayaFranzKeldyshEffectStrongfield2019a} is expected to occur.

The ``in'' creation/annihilation operators are not subjected to the perturbation because they are extracted from the asymptotic field $\Psi^{(0)}$, which is related to the field $\Psi$ through $\Psi \to \sqrt{Z} \Psi^{(0)}$ in the limit $t \to -\infty$. It means that the ``in'' vacuum $\ket{0}_\In$ in \eqref{eq::in_vac} is unperturbed by the additional electric field. The ``out'' creation/annihilation operators, on the other hand, should be extracted from another asymptotic field, denoted by $\Psi_\Out$ with a relation $\Psi \to \sqrt{Z} \Psi_\Out$ in the limit $t \to \infty$, which is dressed by the Furry-picture perturbation. For instance, the ``out'' annihilation operator of particle leaving from the center in the left of the step is defined in the picture (A) as 
\begin{align} 
  a_{L, \Out} (p, s) = \lim_{t \to \infty} \frac{1}{\sqrt{Z}} \int_{-\infty}^\infty dz u_{-p, s}^\dagger (z, t) \Psi (z, t), 
\end{align} 
where 
\begin{align} 
  u_{-p, s} (z, t) = \frac{1}{\sqrt{2\pi}} \sqrt{\frac{m}{E_p}} u(-p, s) e^{-iE_p t - ipz}. \label{eq::u_-p,s} 
\end{align} 
The anti-commutation relations for the ``out'' creation/annihilation operators are guaranteed by assuming the equal-time canonical anti-commutation relations for the asymptotic field $\Psi_\Out$. By the substitution of \eqref{eq::Psi_series}, it is decomposed to the series in terms of $A_3$, i.e., $a_{L, \Out} (p, s) = \sum_{n \geq 0} a_{L, \Out}^{(n)} (p, s)$. In the energy region (i), the zeroth-order term is calculated in Appendix~\ref{AppendB} to be a linear combination of the ``in'' annihilation operators of particles. The limit evaluation yields the next order as 
\begin{align} 
  a_{L, \Out}^{(1)} (p, s) = ie \sqrt{\frac{p}{q} \frac{E_q}{E_p}} \int_{-\infty}^\infty dz dt \bar{\chi}_s^{(V_0 + E_q)} (z, t) \gamma^3 A_3 (z, t) \Psi^{(0)} (z, t), \label{eq::a_Lout(1)} 
\end{align} 
where the transformation for $\chi_s^{(E)}$ in \eqref{eq::lineartrans_modefunct} is used. $\chi_s^{(V_0 + E_q)}$ with the square-root factor in the right-hand side of \eqref{eq::a_Lout(1)} is equivalent to the ``out'' mode function in (A) which characterizes a particle leaving from the center on the left of the step (see \eqref{eq::psi_out}). Observe that the integration over $t$ gives delta functions of energy, implying the energy conservation relations. In particular, the energy eigenvalues of the ``in'' mode functions belonging to the negative-frequency spectra (iii) and (iv) can reach the energy eigenvalue $E_p = V_0 + E_q$ of the ``out'' mode functions $\sqrt{pE_q/qE_p} \, \chi_s^{(V_0 + E_q)}$ with an energy assistance $\omega$ of the perturbative electric field, which leads to the particle--anti-particle mode mixing. For the ``in'' mode functions $\phi_{s'}^{(V_0 - E_{q'})}$, for example, the energy conservation relation is 
\begin{align} 
  E_p = V_0 - E_{q'} + \omega, \label{eq::AssistedEnergyBalance1} 
\end{align} 
implying the energy balance between the particle--anti-particle pair and the external fields (compare it with \eqref{eq::EnergyBalance_step} under the potential step alone). The same relation can be derived for $E_p \leq V_0 + m$. Therefore, it holds for any $p > 0$. Note that for given $p$, the existence of $q'$ satisfying \eqref{eq::AssistedEnergyBalance1} requires a condition 
\begin{align} 
  V_0 + \omega - E_p > m. \label{eq::ExistenceCondition} 
\end{align} 
When we denote its solution as $q' = q_1 (> 0)$, the delta function of the energy is rewritten as 
\begin{align} 
  \delta (E_p - V_0 + E_{q'} - \omega) = \frac{E_{q_1}}{q_1} \delta (q' - q_1), 
\end{align} 
and $\delta (q' - q_1)$ cancels the integration over $q'$ in the field operator $\Psi^{(0)}$. The criticality condition for the vacuum instability is derived from the existence condition \eqref{eq::ExistenceCondition} as 
\begin{align} 
  V_0 + \omega > 2m. \label{eq::CriticalityCondition} 
\end{align} 
This indicates that the assisted pair production can occur even when the potential energy $V_0$ as well as the assistance energy $\omega$ are below the threshold $2m$. Under the condition \eqref{eq::CriticalityCondition}, the number of created particles moving to the left per unit momentum in the distant future 
\begin{align} 
  \Big\langle \frac{dN}{dp} \Big\rangle = \tensor[_\In]{\braket{0| a_{L, \Out}^\dagger (p, s) a_{L, \Out} (p, s) |0}}{_\In} = \tensor[_\In]{\braket{0| a_{L, \Out}^{(1) \dagger} (p, s) a_{L, \Out}^{(1)} (p, s) |0}}{_\In} + \mathcal{O} \bigl( (A_3)^3 \bigr) \label{eq::ParticleNumber_def} 
\end{align} 
becomes finite at the second order of the perturbation. Zeroth- and first-order contributions do not exist because $a_{L, \Out}^{(0)} (p, s)$ does not include ``in'' creation/annihilation operators of anti-particle. One can also calculate the quantity up to the same order for right-moving particles.

Particle creation under the perturbation within the picture (B) is also discussed. A creation/annihilation operator of a particle moving left on the left of the step, for example, is introduced by an analogy with \eqref{eq::c_Lin(0)_def}, as 
\begin{align} 
  c_{L, \Out} (p, s) = \lim_{t \to \infty} \frac{1}{\sqrt{Z}} \int_{-\infty}^\infty dz \varphi_{s, \Out}^{(E_p) \dagger} (z, t) \Psi (z, t), 
\end{align} 
where the ``out'' mode function $\varphi_{s, \Out}^{(E_p)}$ in \eqref{eq::modeft_AB} is used. One can calculate the right-hand side of the above equation in the same way and understand that the first-order term of the perturbation, which causes the particle--anti-particle mode mixing, is different from the ``out'' annihilation operator $a_{L, \Out}^{(1)} (p, s)$ in \eqref{eq::a_Lout(1)}. Therefore, the ``out'' vacuum in the picture (B) differs from that of (A). Since the ``in'' vacuum is unchanged by the perturbation (as explained in the previous paragraph), a differential particle number in the picture (B) is evaluated as 
\begin{align} 
  \tensor[_\In]{\braket{0| c_{L, \Out}^\dagger (p, s) c_{L, \Out} (p, s) |0}}{_\In}. 
\end{align} 
It has a nontrivial value at the second-order perturbation and, interestingly, shows a different momentum distribution from that of \eqref{eq::ParticleNumber_def}, as will be shown in Sec.~\ref{Sec5}.

\section{\label{Sec4}Dynamically assisted pair production}

We investigate the momentum-dependence of the differential particle number within the particle--anti-particle picture (A) up to the second order. A differential number density of left-moving particle, denoted by $n(-p, s)$, is introduced from the differential particle number \eqref{eq::ParticleNumber_def} divided by a volume factor, i.e., 
\begin{align} 
  \Big\langle \frac{dN}{dp} \Big\rangle = n(-p, s) \delta (p - p), 
\end{align} 
where the delta function $\delta (p - p)$ is proportional to an infinite spatial length along $z$-direction. This is brought about from the factors $\tensor[_\In]{\braket{0| bb^\dagger |0}}{_\In}$, where the operators $b, b^\dagger$, written in a shorthand notation, are included in the ``out'' creation/annihilation operators \eqref{eq::a_Lout(1)}. By denoting the mode functions as separated forms to space and time, such as $\psi_s^{(E)} (z, t) = \psi_s^{(E)} (z) e^{-iEt}$, the number density for $E_p = V_0 + E_q > V_0 + m$ is analytically expressed as 
\begin{equation}\begin{split} 
  n(-p, s) = \Bigl( \frac{e\mathcal{E}_z \pi}{\omega} \Bigr)^2 \frac{E_q}{q} &\biggl\{ \theta (E_{q_1} - m) \frac{E_{q_1}}{q_1} \biggl| \sum_{s'} \int_{-\infty}^\infty dz \bar{\chi}_s^{(V_0 + E_q)} (z) \gamma^3 \phi_{s'}^{(V_0 - E_{q_1})} (z) e^{-(z/l)^2} \biggr|^2 \\ 
  &+ \theta (E_{p_2} - m) \frac{E_{p_2}}{p_2} \biggl| \sum_{s'} \int_{-\infty}^\infty dz \bar{\chi}_s^{(V_0 + E_q)} (z) \gamma^3 \psi_{s'}^{(-E_{p_2})} (z) e^{-(z/l)^2} \biggr|^2 \biggr\} + \mathcal{O} (\mathcal{E}_z^3), \label{eq::n(-p,s)} 
\end{split}\end{equation} 
where $q_1 > 0$ is the solution $q' = q_1$ of the equation \eqref{eq::AssistedEnergyBalance1}, and $p_2 > 0$ is determined by another equation of $p'$: 
\begin{align} 
  E_p = -E_{p'} + \omega. \label{eq::AssistedEnergyBalance2} 
\end{align} 
The step functions in \eqref{eq::n(-p,s)} stem from the exsistence conditions for $q_1$ and $p_2$, and the momenta suffice $E_{p_2} = V_0 + E_{q_1}$ if both of them exist, implying $p_1 = p_2$ and $q_1 = q_2$. The first term in the curly brackets of \eqref{eq::n(-p,s)} is a contribution from $\phi_{s'}^{(V_0 - E_{q_1})}$, i.e., negative-frequency electrons on the right of the step, while the second term is one from negative-frequency electrons on the left of the step. By substituting the functional forms of the mode functions, one can perform the integrations over $z$ to obtain expressions with the error function and find that the spin dependence does not appear. For $E_p \leq V_0 + m$, the result \eqref{eq::n(-p,s)} is modified by a following replacement: 
\begin{equation}\begin{split} 
  \sqrt{\frac{E_q}{q}} \chi_s^{(V_0 + E_q)} (z) \xrightarrow{E_p \leq V_0 + m} \sqrt{\frac{E_p}{p}} \psi_s^{(E_p)} (z). \label{eq::continuation_chi_psi} 
\end{split}\end{equation} 
$\psi_s^{(E_p)}$ on the right-hand side of \eqref{eq::continuation_chi_psi} is the ``out'' mode functions (A) of a particle moving to left in the energy region (ii) up to the normalization. Observe that the above replacement of the functions is continuous in terms of the energy eigenvalue because the left-hand side is written, with the aid of the transformation among the mode functions \eqref{eq::lineartrans_modefunct} and the reciprocity \eqref{eq::reciprocity}, as 
\begin{align} 
  \sqrt{\frac{E_p}{p}} \Bigl( R_\psi (p) \psi_s^{(E_p)} (z, t) + \sqrt{\frac{E_q}{E_p}} T_\psi (p) \phi_s^{(V_0 + E_q)} (z, t) \Bigr). 
\end{align} 
At the energy eigenvalue $E_p = V_0 + m$, where the scattering behavior is switched from the over-the-barrier scattering to the total reflection, the reflection and transmission coefficients become $R_\psi (p) = 1$ and $T_\psi (p) = 0$. It assures that $n(-p, s)$ is also continuous in terms of the energy eigenvalue $E_p$, and therefore, the momentum $p$. The differential number density of a right-moving particle, which is defined by factoring $\delta (q - q)$ out of the differential particle number with momentum $q$, is expressed as 
\begin{equation}\begin{split} 
  n(q, s) = \Bigl( \frac{e\mathcal{E}_z \pi}{\omega} \Bigr)^2 \frac{E_p}{p} &\biggl\{ \theta (E_{q_1} - m) \frac{E_{q_1}}{q_1} \biggl| \sum_{s'} \int_{-\infty}^\infty dz \bar{\varphi}_s^{(E_p)} (z) \gamma^3 \phi_{s'}^{(V_0 - E_{q_1})} (z) e^{-(z/l)^2} \biggr|^2 \\ 
  &+ \theta (E_{p_2} - m) \frac{E_{p_2}}{p_2} \biggl| \sum_{s'} \int_{-\infty}^\infty dz \bar{\varphi}_s^{(E_p)} (z) \gamma^3 \psi_{s'}^{(-E_{p_2})} (z) e^{-(z/l)^2} \biggr|^2 \biggr\} + \mathcal{O} (\mathcal{E}_z^3), \label{eq::n(q,s)} 
\end{split}\end{equation} 
with the same notation $q_1$ and $p_2$. Note that the volume factors $\delta (p - p)$ and $\delta (q - q)$ are the same quantity since the momenta in the left and right of the step are written by $\pm p$ and $\pm q$, respectively. Note also that $\varphi_s^{(E_p)}$ on the right-hand side of \eqref{eq::n(q,s)} is proportional to the ``out'' mode function of a right-moving particle, see \eqref{eq::psi_out}.

The left panel of Fig.~\ref{fig::n(k,s)_subcritical} shows the momentum distribution of $n(k, s)$, with $k = -p$ or $q$, when the energy of the oscillating electric field $\omega$ is below the threshold $\omega < 2m$, especially when $\omega = 1.5m$. 
\begin{figure}[htbp] 
  \includegraphics[width=0.45\linewidth]{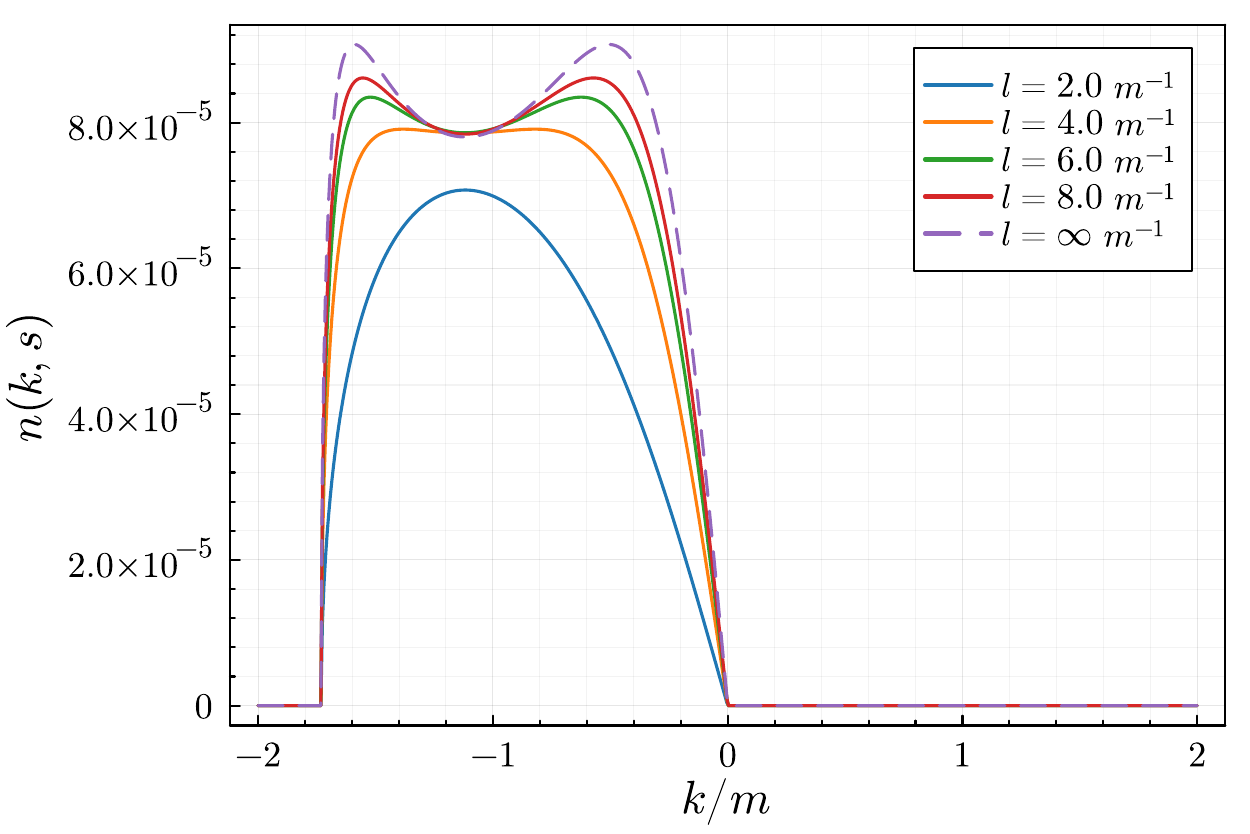} 
  \includegraphics[width=0.4\linewidth]{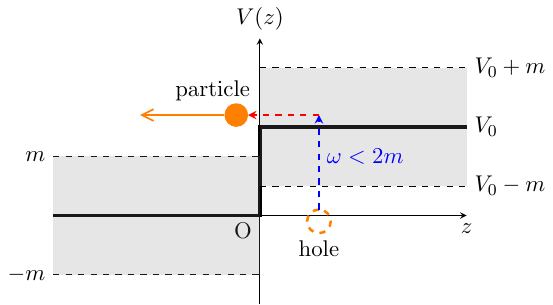} 
  \caption{\label{fig::n(k,s)_subcritical}(Left panel) The momentum distribution of the differential particle number density for $V_0 = 1.5m$ and $\omega = 1.5m$. The other parameters are $\mathcal{E}_z = 0.01m^2/e$ and $l = 2m^{-1}, 4m^{-1}, 6m^{-1}, 8m^{-1}$ and $\infty m^{-1}$. The results for each $l$ are drawn in the blue, orange, green, and red solid lines and purple dashed lines, respectively. (Right panel) The schematic picture of particle production from the Dirac sea when $V_0 < 2m$ and $\omega < 2m$. The created particle moving to the left is drawn in the orange ball with the orange arrow, while the hole in the Dirac sea corresponds to the dashed orange circle. The particle in the site of the hole jumps up and tunnels to the positive-frequency area in $z < 0$ along the blue and red dashed arrows.} 
\end{figure} 
The height of the potential step is also set to the subcritical value $V_0 = 1.5m$, although the criticality condition \eqref{eq::CriticalityCondition} is satisfied. For this case, only $n(-p, s)$ for $E_p \leq V_0 + m$ can be nonvanishing. The field-strength of the oscillating electric field is $\mathcal{E}_z = 0.01m^2/e$, and its spatial width varies from $l = 2m^{-1}$ to $\infty m^{-1}$. The figure shows that the pair production from the vacuum can occur even when the two external fields are insufficient for the vacuum decay. Its mechanism can be understood by looking at the right panel of Fig.~\ref{fig::n(k,s)_subcritical}, where electrons in the Dirac sea in $z > 0$ are excited with the help of $\omega$ and penetrate to the positive-frequency area in $z < 0$, yielding the electron-hole pairs. This is the combined process of the perturbative and non-perturbative contributions, which has already been explained in the original paper of the dynamically assisted Schwinger effect~\cite{schutzholdDynamicallyAssistedSchwinger2008}. It is worth mentioning that the dynamically assisted Schwinger effect under time-dependent backgrounds has been known as the drastic enhancement of the usual Schwinger pair production in which the produced number is exponentially suppressed but nonvanishing. On the other hand, the result obtained here states that the produced number changes from completely zero to a finite value. The difference comes from whether the energy is conserved or not; the system under the potential step has the time-translational invariance, and the energy conservation relations \eqref{eq::AssistedEnergyBalance1} and \eqref{eq::AssistedEnergyBalance2} exactly hold. These relations determine the finite support of the energy distribution $m < E_p < V_0 + \omega - m$. The momentum distribution lies only on the negative region since $E_q = E_p - V_0 > m$ cannot be satisfied for any $q$, or physically speaking, the created particles have no choice but to go left because of the forbidden region in $z > 0$. One can also see that the differential number density increases for a wider area of dynamical assistance $l$ and is maximized for $l = \infty m^{-1}$. The dependence on the parameter is most sensitive at $k \sim -0.5m$ and $k \sim -1.6m$, where two peaks are growing with the increase of $l$. At $k \sim -1.1m$, on the other hand, this dependence hardly exists except for $l = 2m^{-1}$.

The differential number density shows different behavior when the assistance energy exceeds the threshold. The left panel of Fig.~\ref{fig::n(k,s)_overcritical} shows the case $\omega = 2.5m$ with the other parameters the same as those for the previous one. 
\begin{figure}[htbp] 
  \includegraphics[width=0.45\linewidth]{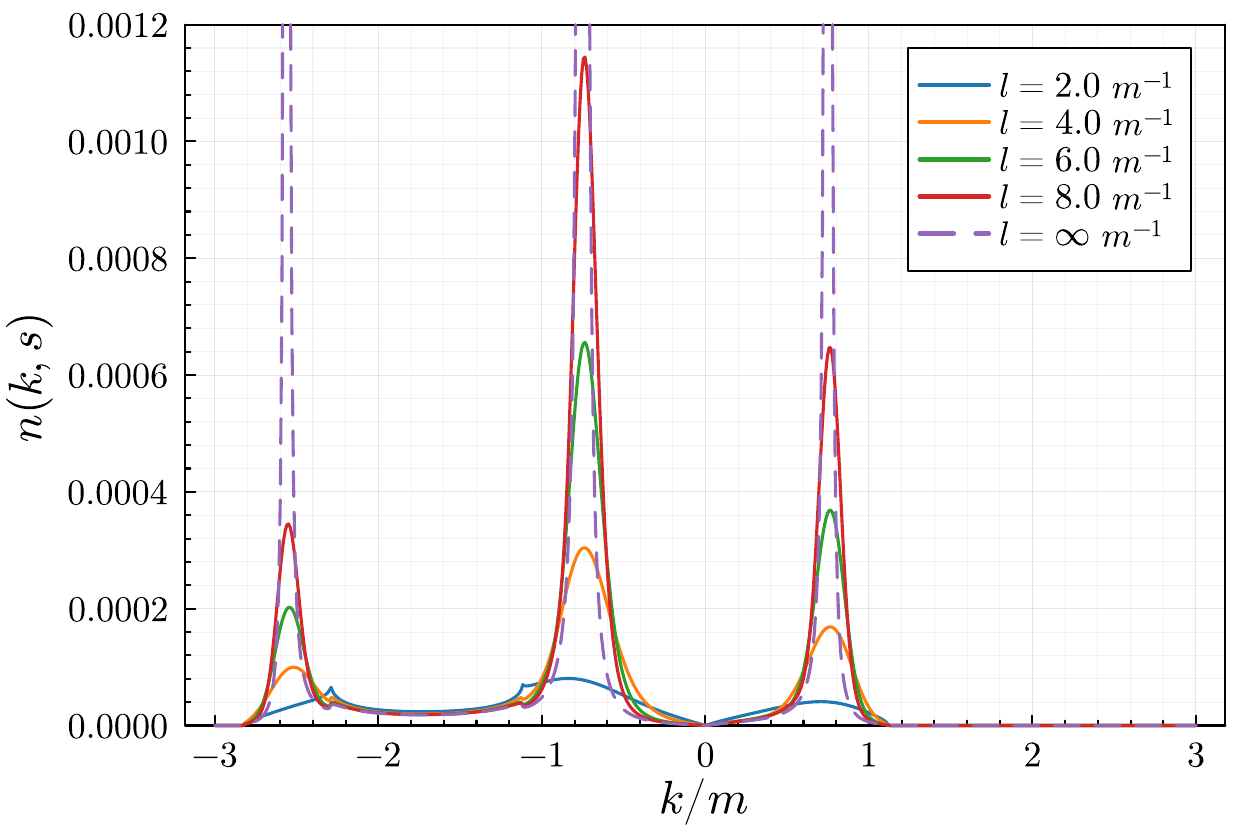} 
  \includegraphics[width=0.4\linewidth]{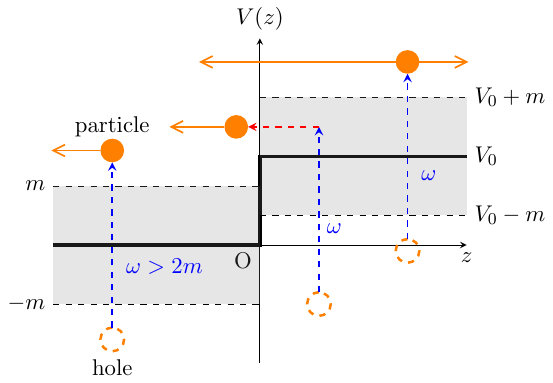} 
  \caption{\label{fig::n(k,s)_overcritical}(Left panel) The momentum distribution of the particle number $V_0 = 1.5m$ and $\omega = 2.5m$. The other parameters are the same as Fig.~\ref{fig::n(k,s)_subcritical}. (Right panel) The schematic picture of particle production from the Dirac sea when $V_0 < 2m$ and $2m < \omega < V_0 + 2m$.} 
\end{figure} 
Now, three peaks are evident for $l \geq 4m^{-1}$, and especially the graph for $l = \infty m^{-1}$ diverges, implying the failure of the perturbative evaluation. The behavior may attribute to the simply perturbative pair production under the oscillating electric field alone~\cite{tayaFinitePulseEffects2014, tayaFranzKeldyshEffectStrongfield2019a} (see also the text~\cite{itzyksonQuantumFieldTheory1980}), where a particle--anti-particle pair can be produced only by the assistance energy $\omega$. However, the perturbative calculations for $l = \infty m^{-1}$ in the previous works give peaks localized in the momentum space by delta functions. The divergences in the current study, on the other hand, is the type of $1/(k - k_0 \pm i\varepsilon)$ ($k_0$ is a divergent point and $\varepsilon$ is a positive infinitesimal), which is brought about from the integration over $z$ in the expressions of the differential number density $n(-p, s)$ and $n(q, s)$. The position $k_0$ of each peak is explained by the particle production described in the right panel of Fig.~\ref{fig::n(k,s)_overcritical}: the simple excitation in the left end contributes to the peak in the middle of the momentum distribution. The combined process of excitation and tunneling can also occur in the present case (drawn in the middle of the three processes), making the middle peak the highest. Particles created by the simple excitation process in the right end of the figure can move toward both directions, generating the peaks in the left and right ends, respectively. Specifically, particles created on the right of the step with negative momenta gain the extra kinetic energy $V_0$ when passing through $z = 0$, which brings about the left-most peak in the far left. For the assistance energy under consideration, the particle created on the left of the step cannot transmit over the potential barrier because $\omega < V_0 + 2m$. If $\omega$ surpasses the threshold, however, the transmission of the particle by losing its energy by $V_0$ can occur, and a new peak accompanied by the process appears in the positive-momentum region. Note that two points $k \sim -1.1m$ and $k \sim -2.3m$, at which the momentum distribution seems not smooth, are found. They correspond to the energy $E_p = V_0$ and $E_p = V_0 + m$ in $n(-p, s)$, respectively, where the mode functions in \eqref{eq::continuation_chi_psi} are continuous but not smooth in terms of the energy eigenvalue. Note also that the two peaks which have been mentioned on the left-hand side of Fig.~\ref{fig::n(k,s)_subcritical} are related to those in the negative-momentum region of the distribution in Fig.~\ref{fig::n(k,s)_overcritical}. The peaks of the differential number density grow with the increase of the assistance energy $1.5m < \omega < 2m$, shifting their positions in the negative direction. When $\omega$ arrives at the threshold $2m$, the peaks become highest at $p = 0$ and $E_p = V_0 + m$, especially those for $l = \infty m^{-1}$ get divergent. Above the threshold energy, the peaks except for $l = \infty m^{-1}$ get lower with shifting positions (to the left for those in the negative-momentum region and to the right for those in the positive-momentum region).

\section{\label{Sec5}Influence of different particle--anti-particle pictures on pair production}

In this section, we evaluate the differential particle number based on the particle--anti-particle picture (B) and compare it with that of (A). The differential number density is introduced by factoring out the delta function of momenta, such as 
\begin{align} 
  \tensor[_\In]{\braket{0| c_{L, \Out}^\dagger (p, s) c_{L, \Out} (p, s) |0}}{_\In} = n'(-p, s) \delta (p - p). 
\end{align} 
The differential number density of the created particle with negative momenta is as follows: for $E_p > V_0 + m$, 
\begin{equation}\begin{split} 
  n'(-p, s) = \Bigl( \frac{e\mathcal{E}_z \pi}{\omega} \Bigr)^2 \frac{E_q}{q} &\biggl\{ \theta (E_{q_1} - m) \frac{E_{q_1}}{q_1} \biggl| \sum_{s'} \int_{-\infty}^\infty dz \bar{\phi}_s^{(V_0 + E_q)} (z) \gamma^3 \chi_{s'}^{(V_0 - E_{q_1})} (z) e^{-(z/l)^2} \biggr|^2 \\ 
  &+ \theta (E_{p_2} - m) \frac{E_{p_2}}{p_2} \biggl| \sum_{s'} \int_{-\infty}^\infty dz \bar{\phi}_s^{(V_0 + E_q)} (z) \gamma^3 \varphi_{s'}^{(-E_{p_2})} (z) e^{-(z/l)^2} \biggr|^2 \biggr\} + \mathcal{O} (\mathcal{E}_z^3), \label{eq::n'(-p,s)} 
\end{split}\end{equation} 
and for $E_p \leq V_0 + m$, the above formula with the replacement 
\begin{align} 
  \sqrt{\frac{E_q}{q}} \phi_s^{(V_0 + E_q)} (z) \xrightarrow{E_p \leq V_0 + m} \sqrt{\frac{E_p}{p}} \psi_s^{(E_p)} (z). \label{eq::replace_phi} 
\end{align} 
In fact, the equality $n'(-p, s) = n(-p, s)$ holds for $E_p \leq V_0 + m$ and an arbitrary $\omega$. This is obvious for $\omega \leq E_p + m$, since $\chi_{s'}^{(V_0 - E_{q_1})}$ in the first line of \eqref{eq::n'(-p,s)} is the same as $\phi_{s'}^{(V_0 - E_{q_1})}$ up to an overall phase factor. It implies that any differences between the differential number density within the two pictures cannot be found for subcritical assistance energy. For $\omega > E_p + m$, the equality is ensured by the following relation 
\begin{equation}\begin{split} 
  &\frac{E_{q_1}}{q_1} \biggl| \sum_{s'} \int_{-\infty}^\infty dz \bar{\psi}_s^{(E_p)} (z) \gamma^3 \chi_{s'}^{(V_0 - E_{q_1})} (z) e^{-(z/l)^2} \biggr|^2 + \frac{E_{p_1}}{p_1} \biggl| \sum_{s'} \int_{-\infty}^\infty dz \bar{\psi}_s^{(E_p)} (z) \gamma^3 \varphi_{s'}^{(-E_{p_1})} (z) e^{-(z/l)^2} \biggr|^2 \\ 
  &= \frac{E_{q_1}}{q_1} \biggl| \sum_{s'} \int_{-\infty}^\infty dz \bar{\psi}_s^{(E_p)} (z) \gamma^3 \phi_{s'}^{(V_0 - E_{q_1})} (z) e^{-(z/l)^2} \biggr|^2 + \frac{E_{p_1}}{p_1} \biggl| \sum_{s'} \int_{-\infty}^\infty dz \bar{\psi}_s^{(E_p)} (z) \gamma^3 \psi_{s'}^{(-E_{p_1})} (z) e^{-(z/l)^2} \biggr|^2. 
\end{split}\end{equation} 
Its validity can be easily checked using the transformation formula among the mode functions \eqref{eq::lineartrans_modefunct}. The differential number density with positive momenta $n'(q, s)$ is also calculated to be 
\begin{equation}\begin{split} 
  n'(q, s) = \Bigl( \frac{e\mathcal{E}_z \pi}{\omega} \Bigr)^2 \frac{E_p}{p} &\biggl\{ \theta (E_{q_1} - m) \frac{E_{q_1}}{q_1} \biggl| \sum_{s'} \int_{-\infty}^\infty dz \bar{\psi}_s^{(E_p)} (z) \gamma^3 \chi_{s'}^{(V_0 - E_{q_1})} (z) e^{-(z/l)^2} \biggr|^2 \\ 
  &+ \theta (E_{p_2} - m) \frac{E_{p_2}}{p_2} \biggl| \sum_{s'} \int_{-\infty}^\infty dz \bar{\psi}_s^{(E_p)} (z) \gamma^3 \varphi_{s'}^{(-E_{p_2})} (z) e^{-(z/l)^2} \biggr|^2 \biggr\} + \mathcal{O} (\mathcal{E}_z^3). 
\end{split}\end{equation} 
The spin dependence does not appear in this particle--anti-particle picture.

\begin{figure}[htbp] 
  \includegraphics[width=0.45\linewidth]{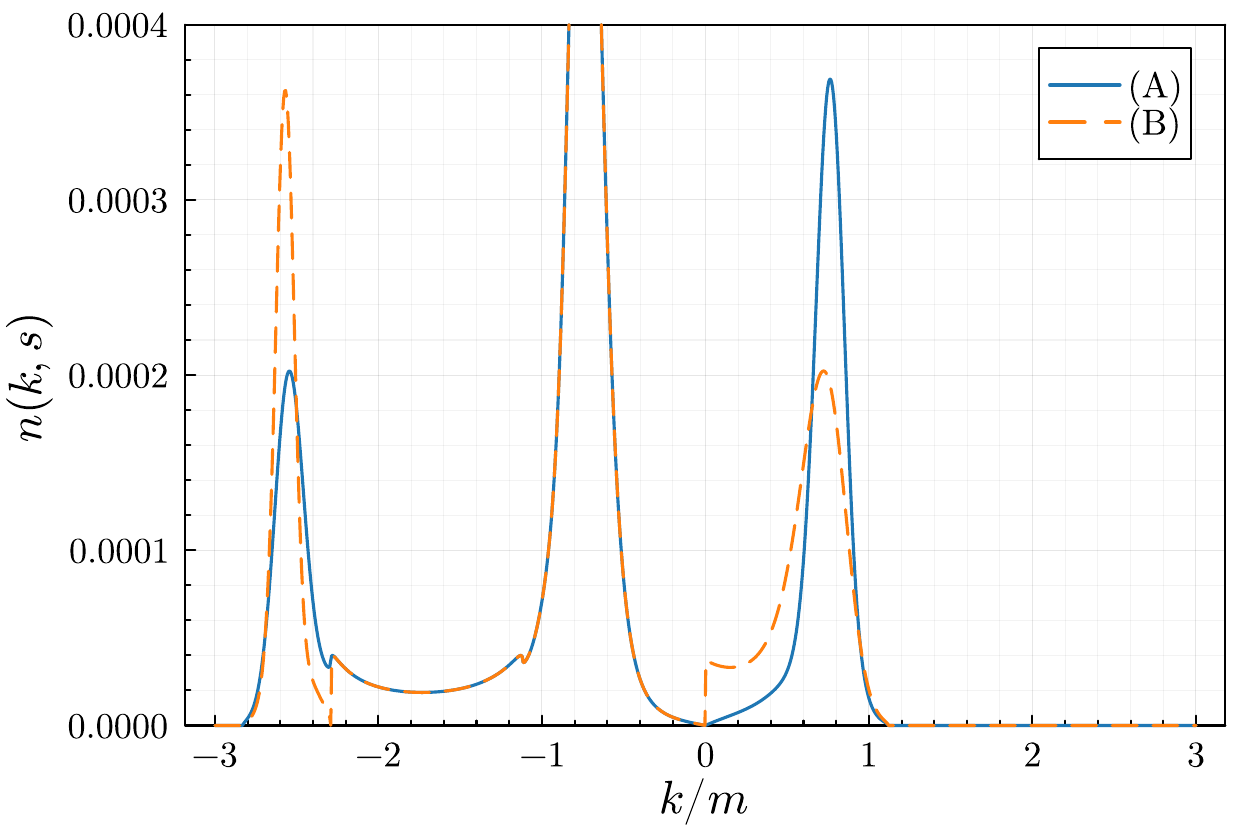} 
  \caption{\label{fig::ab_comparison}Comparison of the momentum distributions of the number density of particles within the particle--anti-particle picture (A) and (B), each of which is expressed in the blue solid line and orange dashed line, respectively. The parameters are $V_0 = 1.5m$, $\omega = 2.5m$, $l = 6m^{-1}$, and $\mathcal{E}_z = 0.01m^2/e$.} 
\end{figure}

Figure~\ref{fig::ab_comparison} shows the comparison of the momentum distributions of the differential number density between the pictures (A) and (B). 
The spatial width of the oscillating electric field is fixed to $l = 6m^{-1}$. The other parameters are the same as Fig.~\ref{fig::n(k,s)_overcritical} (especially the graph expressed in the blue solid line is the same as that in the green solid line in the last figure). The region between $k \sim -2.3m$ and $k = 0$ (corresponding to $E_p = V_0 + m$ and $p = 0$ in $n(-p, s)$, respectively) shows the equivalence of the number densities in (A) and (B) as explained above. The other region exhibits their quantitative difference in that the heights of the leftmost and rightmost peaks are reversed. It may remind the relation, which has been seen in Sec.~\ref{Sec2.2}, that whether a particle is on the left or right of the step is oppositely interpreted in the pictures (A) and (B). Another feature is that the momentum distribution for (B) is discontinuous at $E_p = V_0 + m$ and $p = 0$. It is confirmed analytically from \eqref{eq::n'(-p,s)}. Let us observe its behavior at $E_p = V_0 + m$ for instance: we only need to evaluate the limit $q \to 0$ for the first line of \eqref{eq::n'(-p,s)}, i.e., 
\begin{align} 
  \lim_{q \to 0} \sqrt{\frac{E_q}{q}} \int_{-\infty}^\infty dz \bar{\phi}_s^{(V_0 + E_q)} (z) \gamma^3 \chi_{s'}^{(V_0 - E_{q_1})} (z) e^{-(z/l)^2}. \label{eq::limit_firstline_n'(-p,s)} 
\end{align} 
Since the transmission coefficient in $\phi_s^{(V_0 + E_q)}$ is $T_\phi (q) = \mathcal{O} (q)$ for an infinitesimal $q$, the integration over negative $z$ included in \eqref{eq::limit_firstline_n'(-p,s)} does not give any contribution in the limit $q \to 0$. For positive $z$, the explicit form of $\phi_s^{(V_0 + E_q)}$ is 
\begin{align} 
  \sqrt{\frac{E_q}{q}} \phi_s^{(V_0 + E_q)} (z) = \frac{1}{\sqrt{2\pi}} \sqrt{\frac{m}{q}} \bigl[ u(-q, s) e^{-iy} + R_\phi (q) u(q, s) e^{iy} \bigr], \label{eq::phi_positive_z} 
\end{align} 
where $y = qz$ is introduced as a new integration variable instead of $z$. Notice that the Gaussian convergence factor multiplied by the Jacobian $dz/dy$ produces a delta function in the limit $q \to 0$: 
\begin{align} 
  \frac{1}{q} e^{-(y/(ql))^2} \xrightarrow{q \to 0} l\sqrt{\pi} \delta (y). 
\end{align} 
Thus, the integralation \eqref{eq::limit_firstline_n'(-p,s)} reduces to 
\begin{align} 
  \lim_{q \to 0} l\sqrt{\frac{\pi E_q}{q}} \bar{\phi}_s^{(V_0 + E_q)} (0) \gamma^3 \chi_{s'}^{(V_0 - E_{q_1})} (0). 
\end{align} 
It is easy to check that \eqref{eq::phi_positive_z} at $z = 0$ vanishes for $q \to 0$ because the reflection coefficient is $R_\phi (q) = -1 + \mathcal{O} (q)$, and therefore the differential number density \eqref{eq::n'(-p,s)} at $E_p = V_0 + m$ gives zero for arbitrary $\omega$. The discontinuity of the momentum distribution of the differential number density may brought about from that of the mode functions in terms of their energy eigenvalues. It should be noted that the characteristics shown in this and the previous sections are valid up to the second-order perturbation of the oscillating electric field. The possibility that the higher-order perturbations compensate for the differences in the results between the two pictures cannot be excluded from the analysis.

\section{\label{Sec6}Conclusion and Future work}

In this paper, we discussed the pair production from the vacuum under the subcritical potential step \eqref{eq::V(z)} superimposed by the oscillating electric field \eqref{eq::A_3(z,t)}. One of the results is that vacuum decay by producing particle--anti-particle pairs can occur at the second-order perturbation of the oscillating field when the total energy supplied by the external fields $V_0 + \omega$ is larger than twice the mass $2m$, as shown in Sec.~\ref{Sec4}. The phenomenon can occur especially when both the potential height $V_0$ and the energy carried by the oscillating field $\omega$ are smaller than the threshold energy $2m$, which is nothing but the consequence of the combined effect of the perturbative and non-perturbative pair-creation processes, i.e., the dynamically assisted Schwinger effect. Its underlying mechanism is naively understood with the conventional hole theory that particles in the Dirac sea are excited by the oscillating field and leak out to the positive-frequency area; in that sense, this result is not at all surprising. Note, however, that the criticality condition \eqref{eq::CriticalityCondition} is the exact condition for the vacuum instability and the pair production, i.e., the dynamical assistance makes the particle number finite from the zero. It is in contrast to the case of time-dependent backgrounds, where the vacuum decay rate or the particle number under the strong field alone is exponentially suppressed but remains finite. The difference comes from the energy conservation for the systems under consideration, and it is ensured for the current case due to the time-translational invariance of the potential step. When the assistance energy $\omega$ exceeds the threshold $2m$, the perturbative pair production by the simple excitation can also occur, resulting in significant changes in the momentum distribution of the differential number density. The distribution peaks emerge according to the pair-creating scenarios: the number of peaks and their positions in the momentum space are related to the places of particles created and their directions to move. Note that the result is different from the simple perturbative pair production in \cite{tayaFinitePulseEffects2014, tayaFranzKeldyshEffectStrongfield2019a, itzyksonQuantumFieldTheory1980} in terms of their peak structures. Its origin is also considered the interplay between the perturbative and non-perturbative effects.

Another result is about the influence of the particle--anti-particle pictures on physical quantities. In the quantum field theory under external fields, a fundamental but nontrivial problem of adequately defining particle and anti-particle as well as vacuum arises, especially for the external fields with spatial inhomogeneity. The two previous works~\cite{gavrilovQuantizationChargedFields2016, nakazatoUnstableVacuumFermion2022}, which we called (A) and (B), have tackled the problem in the context of the Klein paradox with their original definitions of the particle--anti-particle notions to obtain the physical quantities such as the vacuum decay rate, the number of created pairs, etc. In the presence of a one-dimensional strong electric field alone, those works bring the same results for the quantities despite the difference of the particle--anti-particle pictures, as explained in Sec.~\ref{Sec2}. We have demonstrated in Sec.~\ref{Sec5} that the differential number density of the created particles exhibits different momentum dependence between the pictures (A) and (B) at the second-order perturbation of the oscillating electric field for $\omega > 2m$. Of course, possibilities that neither of particle--anti-particle interpretations (A) nor (B) are correct still remain. However, the result obtained here implies the dependence on the definition of particles and anti-particles, and it will motivate further discussions by the numerical simulations and experiments in laboratories. An application of this work to some condensed-matter systems such as graphenes~\cite{katsnelsonChiralTunnellingKlein2006, *katsnelsonGrapheneNewBridge2007, *geimGrapheneStatusProspects2009, PhysRevD.78.096009, PhysRevB.81.165431, berdyuginOutofequilibriumCriticalitiesGraphene2022}, which have been of interest as the playgrounds of particle physics without high-energy setups, can be one of the future works. Although we focused only on the particle number in this paper, other physical quantities, such as electric current and scattering amplitudes, are expected to show nontrivial dependence on particle--anti-particle interpretations, which will be reported elsewhere.

Some critical points are left undiscussed. First, the gauge independence of the results is nontrivial since we have fixed the gauge as the background field. One way to address the issue is to replace the scalar potential \eqref{eq::V(z)} with a linear function $A_0 (z) \propto z$, which produces a constant homogeneous electric field. The dynamically assisted pair production under a different gauge, i.e., a linear vector potential $A_3 (t) \propto t$, has been analyzed in \cite{torgrimssonDynamicallyAssistedSauterSchwinger2017, tayaFranzKeldyshEffectStrongfield2019a} using the Furry-picture perturbation theory. Thus, the comparison among them will bring information on the gauge dependence. Another problem is the contributions of higher-order perturbations, as mentioned in the previous section. The differences in the differential particle number density between the two pictures of particle--anti-particle shown in Fig.~\ref{fig::ab_comparison} may be attributed to the truncation of the perturbation expansion. Although we have given the field strength of the oscillating electric field as sufficiently weak compared to the Schwinger limit, the validity of the perturbation theory should be checked numerically. In the numerical simulation, adiabatically switching on/off has to be incorporated into the external fields. Note that the previous works \cite{torgrimssonDynamicallyAssistedSauterSchwinger2017, tayaFranzKeldyshEffectStrongfield2019a} have conducted the numerical simulation of the pair production to show that the particle number density estimated up to the second order is in good agreement with the numerical results. The optimal order of the perturbation will also be of interest in the current case.

\appendix 
\section{\label{AppendA}Dirac spinors}

Dirac spinors and their basic properties are summarized here. A four-component Dirac field $\psi$ of a free relativistic fermion of mass $m$ in the four-dimensional spacetime obeys the free Dirac equation 
\begin{align} 
  (i\slashed{\partial} - m) \psi = 0, \label{eq::freeDiracEq} 
\end{align} 
where $\slashed{\partial} = \gamma^\mu \partial_\mu$ ($\mu = 0, 1, 2, 3$) and $\gamma^\mu$ are $4\times 4$ gamma matrices. An energy eigenvalue $E$ of $\psi$ belongs to two continuous spectra $|E| \geq m$, split by a mass gap between $\pm m$. Neglecting transversal directions to $z$-axis, the positive- and negative-frequency solutions of \eqref{eq::freeDiracEq} for momentum $k$ and spin $s$ are written as 
\begin{align} 
  \frac{1}{\sqrt{2\pi}} \sqrt{\frac{m}{E_k}} u(k, s) e^{-iE_k t + ikz}, \quad \frac{1}{\sqrt{2\pi}} \sqrt{\frac{m}{E_k}} v(k, s) e^{iE_k t - ikz}, \label{eq::planewave_sols} 
\end{align} 
($E_k = \sqrt{k^2 + m^2}$ is the on-shell energy) respectively. The positive- and negative-frequency spinors $u$ and $v$ are expressed in the Dirac representation as 
\begin{align} 
  u(k, s) = \sqrt{\frac{E_k + m}{2m}} \begin{pmatrix} \1 \\ \frac{k \sigma_z}{E_k + m} \end{pmatrix} \bm{\xi} (s), \quad v(k, s) = \sqrt{\frac{E_k + m}{2m}} \begin{pmatrix} \frac{k \sigma_z}{E_k + m} \\ \1 \end{pmatrix} \bm{\xi} (s) \label{eq::u_v} 
\end{align} 
with a two-component spinor $\bm{\xi} (s)$, which is factored out of the four-component spinors. $u$ and $v$ in \eqref{eq::u_v} satisfy the normalization and orthogonality conditions 
\begin{align} 
  u^\dagger (k, s) u(k, s') = \delta_{s, s'} = v^\dagger (k, s) v(k, s'), \quad u^\dagger (k, s) v(-k, s') = 0 
\end{align} 
as well as the completeness relation 
\begin{align} 
  \sum_{s = \pm} \bigl[ u_\alpha (k, s) u_\beta^\dagger (k, s) + v_\alpha (-k, s) v_\beta^\dagger (-k, s) \bigr] = \delta_{\alpha, \beta}, 
\end{align} 
where $\alpha, \beta = 1, 2, 3, 4$ stand for spinor indices. When a constant scalar potential such as $V(z) = V_0$ is imposed on \eqref{eq::freeDiracEq}, an extra phase factor $\exp (-iV_0 t)$ is applied to the plane-wave solutions \eqref{eq::planewave_sols}. Thus, the positive-frequency solutions belong to the energy eigenvalue $E > V_0 + m$, while the negative-frequency solutions belong to $E < V_0 - m$.

\section{\label{AppendB}``Out'' creation/annihilation operators and mode functions within the picture (A)}

Within the particle--anti-particle picture (A), ``out'' annihilation operators of particles in the left and right of the step, under the potential step alone, are defined as 
\begin{align} 
  a_{L, \Out}^{(0)} (p, s) &= \lim_{t \to \infty} \int_{-\infty}^\infty dz u_{-p, s}^\dagger (z, t) \Psi^{(0)} (z, t), \\ 
  a_{R, \Out}^{(0)} (q, s) &= \lim_{t \to \infty} \int_{-\infty}^\infty dz u_{q, s}^\dagger (z, t) \Psi^{(0)} (z, t), \label{eq::a_Rout(0)_def} 
\end{align} 
where $u_{-p, s}$ is the left-moving monochromatic plane wave \eqref{eq::u_-p,s} and $u_{q, s}$ is the right-moving one: 
\begin{align} 
  u_{q, s} (z, t) = \frac{1}{\sqrt{2\pi}} \sqrt{\frac{m}{E_q}} u(q, s) e^{-i(V_0 + E_q) t + iqz}. 
\end{align} 
In \eqref{eq::a_Rout(0)_def}, the outgoing particle in the right of the step is characterized by the positive-frequency plane wave in the parentheses, with the factor $\exp (-iV_0 t)$ mentioned above. The limit evaluation of the right-hand sides of the above equations in the same way as the discussion in \eqref{eq::a_Lin(0)_def} yields the ``in/out'' transformation among the asymptotic creation/annihilation operators: for example, in the energy range (i) $E_p = V_0 + E_q > V_0 + m$, 
\begin{align} 
  \begin{pmatrix} a_{L, \Out}^{(0)} (p, s) \\ a_{R, \Out}^{(0)} (q, s) \end{pmatrix} = \begin{pmatrix} R_\psi (p) & \frac{p}{q} \sqrt{\frac{E_q}{E_p}} T_\phi (q) \\ \frac{q}{p} \sqrt{\frac{E_p}{E_q}} T_\psi (p) & R_\phi (q) \end{pmatrix} \begin{pmatrix} a_{L, \In}^{(0)} (p, s) \\ a_{R, \In}^{(0)} (q, s) \end{pmatrix}. \label{eq::Bogoliubov_trans} 
\end{align} 
The use of the above relation, the probability conservation \eqref{eq::prob_conserv}, and the reciprocity \eqref{eq::reciprocity} confirms the anti-commutation relations for the ``out'' annihilation operators. Note that \eqref{eq::Bogoliubov_trans} is not the unitary transformation because the momenta $p$ and $q$ are not well-defined quantum numbers. However, this is essentially the same as the so-called Bogoliubov transformation in the previous works \cite{gavrilovQuantizationChargedFields2016}, where energy eigenvalues instead of momenta characterize creation/annihilation operators.

The ``out'' mode functions (A) are introduced as the basis functions of the field operator $\Psi^{(0)}$ when it is decomposed to a sum of the ``out'' creation/annihilation operators. The ``out'' mode functions accompanied by $a_{L, \Out}^{(0)} (p, s)$ and $a_{R, \Out}^{(0)} (q, s)$ in the energy region (i), denoted by $\psi_{s, \Out}^{(E)}$ and $\phi_{s, \Out}^{(E)}$, are extracted from the field operator by the following anti-commutators 
\begin{align} 
  \psi_{s, \Out}^{(E)} (z, t) = \{ a_{L, \Out}^{(0) \dagger} (p, s), \Psi^{(0)} (z, t) \}, \quad \phi_{s, \Out}^{(E)} (z, t) = \{ a_{R, \Out}^{(0) \dagger} (q, s), \Psi^{(0)} (z, t) \}. 
\end{align} 
By using \eqref{eq::Bogoliubov_trans} and the field expansion \eqref{eq::field(A)}, they are calculated to give 
\begin{align} 
  \psi_{s, \Out}^{(E)} (z, t) = \sqrt{\frac{p}{q} \frac{E_q}{E_p}} \chi_s^{(E)} (z, t), \quad \phi_{s, \Out}^{(E)} = \sqrt{\frac{q}{p} \frac{E_p}{E_q}} \varphi_s^{(E)} (z, t), \label{eq::psi_out} 
\end{align} 
which means that the outgoing particle (A) on the left(right) of the step is interpreted in the other picture as the incoming particle on the right(left) of the step. The correspondence between the two pictures is also seen for the other ``out'' mode functions in the regions (i) and (iv). Note that the overall factor in the right-hand side of the above equation is determined by the normalization conditions \eqref{eq::norm_psi}, while $\chi_s^{(E)}$ is done by \eqref{eq::norm_phi}.

\begin{acknowledgments}
The author would like to thank K. Yamashiro for the fruitful discussion at the early stage of this work. The author is grateful to H. Nakazato for his critical reading of the manuscript and helpful advice. 
\end{acknowledgments}

\bibliographystyle{apsrev4-2}
\bibliography{refs.bib}

\end{document}